\documentclass[11pt]{article}
\usepackage{amsmath,amssymb}
\usepackage{epsfig,graphics,graphicx}
\usepackage{color,multicol}
\usepackage{natbib}
\usepackage{multirow}
\usepackage{url}

\usepackage{color, colortbl}
\definecolor{Gray}{gray}{0.9}

\begin{document}

\title{Bayesian Mixture Modeling for Multivariate Conditional Distributions}
\author{Maria De Yoreo and Jerome P. Reiter  \thanks{M. DeYoreo (maria.deyoreo@stat.duke.edu) is postdoctoral researcher and J. Reiter is Mrs. Alexander Hehmeyer Professor of Statistics, Department of Statistical Science. This research was supported in part by \textit{The National Science Foundation} under 
award SES-11-31897.}} 
\date{}

\maketitle 

\begin{abstract} 
We present a Bayesian mixture model for estimating the joint distribution of
mixed ordinal, nominal, and continuous data conditional on a set of
fixed variables. The model uses multivariate normal and
categorical mixture kernels for the random variables. It induces
dependence between the random and fixed variables through the means of
the multivariate normal mixture kernels and via a truncated local
Dirichlet process.  The latter encourages observations with
similar values of the fixed variables to share mixture
components. Using a simulation of data fusion, we illustrate that the
model can estimate underlying relationships in the data and the
distributions of the missing values more accurately than a mixture model applied to the random and fixed
variables jointly.  We use the model to analyze consumers' reading
behaviors using a quota sample, i.e., a sample where the empirical
distribution of some variables is fixed by design and so should not be
modeled as random, conducted by the book publisher HarperCollins. 
\end{abstract}


\section{Introduction}\label{intro}

Bayesian mixture models are flexible and convenient tools for estimating the joint distribution of a set of variables \citep[e.g.,][]{dunsonxing,banerjee,muller:mitra}.  Often, 
however, it is desirable to treat some of the variables as conditioning information rather than random variables.  For example, when data are collected using 
a stratified or quota sampling design, the empirical distribution of the design
variables is fixed {\em a priori}.  On principle, it 
does not make sense to estimate their distribution with uncertainty.  In fact, when the sampling is not proportional 
to population shares, treating the design variables as random can result in badly biased estimates of population-level 
quantities \citep{schifeling:reiter, kunihama:herring, fosdick}.  As another example, when using mixture models for multiple imputation 
of missing data \citep{rubin}, it is unnecessary to estimate the marginal distribution of the variables with no missing values. Rather, 
all we need is the conditional distribution of variables with missingness given those that are fully observed.

In this article, we present mixture models for estimating the joint
distribution of variables treated as random conditional on a set of  variables treated as fixed.  The models use  
multivariate normal kernels for continuous and ordinal variables (via a probit specification), and independent multinomial kernels for nominal variables.  We induce dependence 
between the random and fixed variables in two ways.  First, for
continuous and ordinal variables, we let the mean of the multivariate
normal distribution within each mixture component  
be a function of the fixed variables.  Second, we encourage observations that have similar values of the fixed variables to share mixture components via a local Dirichlet process \citep{chung}.
The local Dirichlet process facilitates estimation of the dependence between the
nominal random variables and the fixed variables, which is otherwise difficult to capture.  It also offers the model 
additional flexibility to capture relationships between the random
continuous/ordinal variables and the fixed variables.



The proposed conditional mixture model can have advantages over other approaches commonly used for estimating multivariate conditional distributions of mixed data.  
For example, one alternative is to eschew mixtures altogether and 
specify multivariate linear or logistic regressions. Such models make strong assumptions, e.g., linearity and Gaussian errors, and can require challenging model specification tasks, e.g., 
selecting which interaction effects to include in logistic
regressions.  By comparison, mixture models tend to be more capable at capturing complex distributional features \citep{muller:quintana,Norets,Papageorgiou}.
Another alternative is to treat all variables as random, estimate their joint distribution via a mixture model, and derive relevant conditional distributions from the resulting estimates \citep{muller:erkanli:west,shahbaba:neal,dunson:bhat,hannah}.   
Even with (modest-sized) representative samples, a full mixture model can waste fitting power on 
the joint distribution of the fixed variables, as it seeks to fit the entire joint distribution.  This can result in 
poor predictive inference for conditional distributions
\citep{wade:dunson}.  A third alternative is to use a mixture
model in which the mixture weights depend on the fixed
variables \citep[e.g.,][]{griffin:steel,dunson:park}.
However, it can be complicated to estimate and obtain inferences
from such models, particularly when the number of fixed variables is
not small.  

The remainder of this article is organized as follows. In Section \ref{sec:methodology}, we describe the mixture model for multivariate conditional inference with mixed data, 
which we refer to as CMM-Mix.  We also describe 
a variable selection procedure based on estimated mutual information values \citep{battiti,ding:peng,estevez} that can be used to trim variables from the conditioning set that determines the local weights, which can be useful when 
the set contains many variables.  
In Section \ref{sec:DataFusion}, we compare CMM-Mix to a full mixture
model in a simulation study of techniques for data fusion
\citep{rassler, gilula,dorazio}, which is a type of missing data scenario 
common in marketing contexts.
In Section \ref{sec:HarperCollins}, we analyze data from 
a quota sample from HarperCollins Publishers, in which we seek to understand relationships involving individuals' reading behaviors and interests; for example, what distinguishes people who own eBooks from those who do not? 
In Section \ref{sec:Discussion}, we 
conclude  with future directions for research.  
This article is  accompanied by supplementary material 
that presents additional results from simulations and data illustrations.

\section{Methodology}
\label{sec:methodology}

Suppose that the collected data include $p$ variables that the analyst treats as random, and $q$ variables that the analyst treats as fixed.
For $i=1, \dots, n$, let $Y_{ij}^{(R)}\in \{1,\dots, k_j^{(R)}\}$ be the value of ordinal random variable $j$ for individual $i$, 
for $j=1,\dots,p_o$; let $X_{ij}^{(R)}\in\{1,\dots,d_j^{(R)}\}$ be the value of nominal random variable $j$ for individual $i$, for $j=1,\dots,p_n$; and, let $Z_{ij}^{(R)}$ be the 
standardized value of continuous random variable $j$ for individual $i$, for $j=1,\dots,p_c$. To facilitate modeling, 
we introduce a latent continuous random variable $W_{ij}^{(R)}$ for each $Y_{ij}^{(R)}$. Similarly, for each $i$ let there be $q_o$ ordinal fixed variables $Y_{ij}^{(F)}\in \{1,\dots, k_j^{(F)}\}$, 
$q_n$ nominal fixed variables $X_{ij}^{(F)}\in\{1,\dots,d_j^{(F)}\}$,
and $q_c$ standardized continuous fixed variables $Z_{ij}^{(F)}$.  We
write each individual's data as the vector $(\mathbf{Y}_i^{(R)},\mathbf{Z}_i^{(R)},\mathbf{X}_i^{(R)}, \mathbf{Y}_i^{(F)},\mathbf{Z}_i^{(F)},\mathbf{X}_i^{(F)})$.
Writing the variables generically, we seek to construct a mixture model for
$p(\mathbf{Y}^{(R)},\mathbf{Z}^{(R)},\mathbf{X}^{(R)}\mid
\mathbf{Y}^{(F)},\mathbf{Z}^{(F)},\mathbf{X}^{(F)})$. 
To simplify notation, we 
sometimes refer to the set 
$(\mathbf{Y}^{(R)},\mathbf{Z}^{(R)},\mathbf{X}^{(R)})$ as
$\boldsymbol{\mathcal{R}}$ and the set $(\mathbf{Y}^{(F)},\mathbf{Z}^{(F)},\mathbf{X}^{(F)})$ as 
$\boldsymbol{\mathcal{F}}$. 

\subsection{Modeling strategy: Connecting $\boldsymbol{\mathcal{R}}$
  and $\boldsymbol{\mathcal{F}}$ via CMM-Mix}

Let $H_i\in\{1,\dots,N\}$ be a mixture allocation variable
representing the component observation $i$ belongs to, such that
$H_i=l$ if and only if observation $i$ belongs to component $l$.  At
the level of the data, for any individual $i$ the model for CMM-Mix is  
\begin{eqnarray} \label{eqn:datamodel1} 
(\mathbf{W}_{i}^{(R)},\mathbf{Z}_i^{(R)}\mid
\{\boldsymbol{\beta}_h\},\{\boldsymbol{\Sigma}_h\},H_i,\mathbf{X}^{(R)}_i,\boldsymbol{\mathcal{F}})
&\sim& \mathrm{N}(\boldsymbol{D}(\mathbf{X}^{(R)}_i,\mathbf{Y}_i^{(F)},\mathbf{Z}_i^{(F)},\mathbf{X}_i^{(F)})\boldsymbol{\beta}_{H_i},\boldsymbol{\Sigma}_{H_i}) \\
(\mathbf{X}_{i}^{(R)} \mid \{\boldsymbol{\psi}_h\},H_i) &\sim& \prod_{j=1}^{p_n}\mathrm{categ}(\psi_{H_i,1}^{(j)},\dots,\psi_{H_i,d_j^{(R)}}^{(j)}).\label{eqn:datamodel1a}
\end{eqnarray}
Here, $\boldsymbol{D}(\cdot)$ is a design vector of length $k$
encoding main effects and possibly non-linear terms identified
through exploratory data analysis as helpful for capturing local dependence, and
$\boldsymbol{\beta}_h$ is a $k\times (p_{c}+p_o)$ matrix of regression
coefficients.    The ordinal $\mathbf{Y}^{(R)}$ are determined from
$\mathbf{W}^{(R)}$ such that $Y^{(R)}_{ij}=l$ if and only if
$\gamma_{j,l-1}<W^{(R)}_{ij}\leq\gamma_{j,l}$, for
$l=1,\dots,k_j^{(R)}$, and
$-\infty=\gamma_{j,0}<\gamma_{j,1}<\dots<\gamma_{j,k_j^{(R)}-1}<\gamma_{j,k_j^{(R)}}=\infty$. 
 

The mixture of regressions helps  capture dependence between
$(\mathbf{Y}^{(R)},\mathbf{Z}^{(R)})$ and 
$\boldsymbol{\mathcal{F}}$.  It also strengthens the connections
between $(\mathbf{Y}^{(R)},\mathbf{Z}^{(R)})$ and
$\mathbf{X}^{(R)}$ by allowing for local dependence within
components. This regression approach   
is related to the ANOVA dependent Dirichlet process (DP) model of
\citet{deiorio} and also is used by
\citet{murray} in a full mixture model.  We explicitly avoid the regression approach for
connecting $\mathbf{X}^{(R)}$ and  $\boldsymbol{\mathcal{F}}$,
primarily because it can be 
computationally challenging to implement with MCMC sampling.  For example, 
multinomial logistic regressions for $\mathbf{X}^{(R)}$ within
components can introduce 
a large number of regression parameters for which there are no
conjugate priors.  

With only \eqref{eqn:datamodel1} and \eqref{eqn:datamodel1a}, the
model has to capture dependence between $\mathbf{X}^{(R)}$ and
$\boldsymbol{\mathcal{F}}$ through a convoluted path involving the
regressions for $(\mathbf{Y}^{(R)},\mathbf{Z}^{(R)})$. With modest
sample sizes, this path seems unlikely to be up to the task.  We
therefore strengthen the connections using a truncated local Dirichlet
process (LDP).  This provides a prior distribution for a collection of random distributions indexed by fixed variables $\boldsymbol{\mathcal{F}}$, in which units that have similar values of $\boldsymbol{\mathcal{F}}$ are assumed to share similar distributions for $\boldsymbol{\mathcal{R}}$.  The dependence between distributions associated with fixed variables ${\boldsymbol{f}}$ and ${\boldsymbol{f}}'$ increases as the distance between ${\boldsymbol{f}}$ and ${\boldsymbol{f}}'$ decreases. This prior possesses many attractive properties, such as retaining a marginal DP prior for any $\boldsymbol{f}$, as described by \citet{chung}.
 

Each $H_i$ arises from a subset of $\{1,\dots,N\}$, where the particular subset
is chosen according to the value of $\boldsymbol{\mathcal{F}}$ for
observation $i$.  The truncated LDP is built from sequences of
mutually independent random variables: $\{V_h: h=1,2,\dots, N\}$ are
beta$(1,\alpha)$ distributed random variables that determine the
stick-breaking weights,
$\{\boldsymbol{\theta}_h=(\boldsymbol{\beta}_h,\boldsymbol{\Sigma}_h,\boldsymbol{\psi}_h):h=1,2,\dots,
N\}$ are the atoms, and $\{\boldsymbol{\Gamma}_h:h=1,2,\dots, N\}$ are
locations in $\boldsymbol{\mathcal{S}}$, the sample space of $\boldsymbol{\mathcal{F}}$.
The mixture weights and probabilities associated with an observation
having a particular $\boldsymbol{\mathcal{F}}=\boldsymbol{f}$ are
determined by the set of $\boldsymbol{\Gamma}_h$ that are in some
neighborhood of $\boldsymbol{f}$. Specifically, let
$\eta_{\boldsymbol{f}}=\{h:d(\boldsymbol{f},\boldsymbol{\Gamma}_h)\leq
d^*\}$ be a set indexing the locations belonging to the
$d^*$--neighborhood of $\boldsymbol{f}$. Here $d$ is a distance
measure and $d^*$ represents the neighborhood size. Elements
$\eta_{\boldsymbol{f}}$ of $\{V_h\}$ and $\{\boldsymbol{\theta}_h\}$
are then used in constructing $G_{{\boldsymbol{f}}}$, the random
distribution associated with  $\boldsymbol{f}$. The prior on the mixture components and weights is therefore:
\begin{eqnarray}\label{eqn:hier_ldp}
H_i\mid \boldsymbol{V},\boldsymbol{\Gamma},\boldsymbol{\mathcal{F}}_i=\boldsymbol{f}_i\stackrel{ind.}{\sim} \sum_{h=1}^{N(\boldsymbol{f}_i)}p_h(\boldsymbol{f}_i)\delta_{\pi_h(\boldsymbol{f}_i)}(\cdot), \; i=1,\dots,n \nonumber \\ 
V_h\mid\alpha\stackrel{i.i.d.}{\sim} \mathrm{beta}(1,\alpha), \; h=1,\dots,N \nonumber \\
\boldsymbol{\Gamma}_h\stackrel{i.i.d.}{\sim} p(\boldsymbol{\Gamma}_h), \; h=1,\dots,N 
\end{eqnarray}
where
$p_h(\boldsymbol{{f}}_i)=V_{\pi_h(\boldsymbol{f}_i)}\prod_{j<h}(1-V_{\pi_j(\boldsymbol{f}_i)})$
for $h=1,\dots,N(\boldsymbol{f}_i)-1$, $\pi_h(\boldsymbol{f}_i)$ is
the $h$th ordered index in $\eta_{\boldsymbol{f}_i}$, and
$N(\boldsymbol{f}_i)=|\eta_{\boldsymbol{f}_i}|$. The last element in
each probability vector $p_{N(\boldsymbol{f}_i)}(\boldsymbol{f}_i)$ is
determined so that
$\sum_{h=1}^{N(\boldsymbol{f}_i)}p_h(\boldsymbol{f}_i)=1$. Although a
global truncation level of $N$ is fixed, $N(\boldsymbol{f}_i)$ is
still a random quantity. Thus, the number of components in each
mixture distribution is data driven. For $p(\boldsymbol{\Gamma}_h)$,
we use a product of independent uniforms for each variable in
$\boldsymbol{\mathcal{F}}$. We discuss the choice of distance
function in Section \ref{sec:distance}.

The base distributions from which the atoms $\boldsymbol{\theta}_h$ are drawn from are given by:
\begin{eqnarray}\label{eqn:atoms_prior}
\boldsymbol{\beta}_h \stackrel{i.i.d.}{\sim} \mathrm{MN}_{k\times (p_o+p_c)}(\boldsymbol{\beta}_0,\mathrm{diag}(\tau_1^2,\dots,
\tau_{k}^2),\boldsymbol{I}_{p_o+p_c}), \; h=1,\dots,N \nonumber\\
\boldsymbol{\Sigma}_h\stackrel{i.i.d.}{\sim}  \mathrm{IW}(\nu,\boldsymbol{S}), \; h=1,\dots,N\nonumber \\
\boldsymbol{\psi}_h^{(j)}\stackrel{i.i.d.}{\sim}  \mathrm{Dirichlet}(a_1^{(j)},\dots,a_{d_j}^{(j)}) , \; h=1,\dots,N, j=1,\dots,p_n
\end{eqnarray}
where $\mathrm{MN}_{k\times (p_o+p_{c})}$ denotes a matrix-normal
distribution of dimension $k$ by $p_o+p_{c}$. This implies that
  $\mathrm{vec}(\boldsymbol{\beta}_h)\sim
  \mathrm{N}_{k(p_o+p_{c})}(\mathrm{vec}(\boldsymbol{\beta}_0), \boldsymbol{I}_{p_o+p_c}\otimes
  \mathrm{diag}(\tau_1^2,\dots,\tau_{k}^2))$, where vec($\boldsymbol{\beta}_h$) denotes the
  vectorization of $\boldsymbol{\beta}_h$, obtained by stacking its
  columns. 
The model is completed with hyperpriors on the parameters
$\boldsymbol{\beta}_0$, $\boldsymbol{S}$, $\boldsymbol{\tau}$, and
$\alpha$. Prior specification is discussed in Appendix \ref{sec:prior}, including justification for the choice of the base distribution for  $\boldsymbol{\beta}_h$.

\subsection{Model properties}
\label{sec:properties}


To describe the properties of CMM-Mix, it is useful first to marginalize over the mixture
allocation indicators to obtain the multivariate conditional density
for  $\boldsymbol{\mathcal{R}}$. We have
$f(\mathbf{W}^{(R)},\mathbf{Z}^{(R)},\mathbf{X}^{(R)}\mid \boldsymbol{\mathcal{F}}=\boldsymbol{f}) =$  
\begin{equation}\label{eqn:joint_dens}
\sum_{h=1}^{N(\boldsymbol{\mathcal{F}})} p_h(\boldsymbol{f})  \mathrm{N}(\mathbf{W}^{(R)},\mathbf{Z}^{(R)};  \boldsymbol{D}(\mathbf{X}^{(R)},\boldsymbol{f})\boldsymbol{\beta}_{\pi_h(\boldsymbol{f})},\boldsymbol{\Sigma}_{\pi_h(\boldsymbol{f})})\prod_{j=1}^{p_n}\mathrm{categ}(X_j^{(R)};\boldsymbol{\psi}_{\pi_h(\boldsymbol{f})}^{(j)}).
\end{equation} 
We manipulate this expression to derive statements about the multivariate conditional
distributions of the random variables at any $\boldsymbol{f}$.

Marginalizing \eqref{eqn:joint_dens} over $\mathbf{W}^{(R)}$ and
$\mathbf{Z}^{(R)}$, we find that the marginal distribution for $\mathbf{X}^{(R)}$
is a mixture of independent multinomials. In particular, we have
$\mathrm{Pr}(\mathbf{X}^{(R)}=\mathbf{x}\mid
\boldsymbol{\mathcal{F}}=\boldsymbol{f})=\sum_{l=1}^{N(\boldsymbol{f})}p_{l}(\boldsymbol{f})\prod_{j=1}^{p_n}\psi^{(j)}_{\pi_l(\boldsymbol{f}),x_j}$. Mixtures
of multinomials are quite effective for modeling multivariate categorical data
distributions \citep{dunsonxing}.  As desired, the model can capture  dependencies between
$\mathbf{X}^{(R)}$ and $\boldsymbol{\mathcal{F}}$, as  the weights and mixture component parameters depend on
$\boldsymbol{\mathcal{F}}$.

The 
$f(\mathbf{W}^{(R)},\mathbf{Z}^{(R)}\mid \mathbf{X}^{(R)},\boldsymbol{\mathcal{F}})$ is a mixture of
multivariate normal linear regressions, with means that are functions of $(\mathbf{X}^{(R)}, \boldsymbol{\mathcal{F}})$ and weights that are functions of
$\boldsymbol{\mathcal{F}}$. We have  
\begin{equation}\label{eqn:cont-cond-X}f(\mathbf{W}^{(R)},\mathbf{Z}^{(R)}\mid
\mathbf{X}^{(R)}=\boldsymbol{x},\boldsymbol{\mathcal{F}}=\boldsymbol{f})=\sum_{h=1}^{N(\boldsymbol{f})} p_h(\boldsymbol{f}) \mathrm{N}(\mathbf{W}^{(R)},\mathbf{Z}^{(R)};\boldsymbol{D}(\boldsymbol{x},\boldsymbol{f})\boldsymbol{\beta}_{\pi_h(\boldsymbol{f})},\boldsymbol{\Sigma}_{\pi_h(\boldsymbol{f})}).\end{equation} 
From (\ref{eqn:cont-cond-X}) we can integrate out $\mathbf{Z}^{(R)}$
and obtain the probability that $\mathbf{Y}^{(R)}$ takes on a
particular combination of ordinal levels $(y_1,\dots,y_{p_o})$
conditional on $\boldsymbol{\mathcal{F}}$ and $\mathbf{X}^{(R)}$. This gives $\mathrm{Pr}(\mathbf{Y}^{(R)}= (y_1,\dots,y_{p_o})\mid \mathbf{X}^{(R)}=\boldsymbol{x},\boldsymbol{\mathcal{F}}=\boldsymbol{f})=$  
\begin{equation}\label{eqn:ord-reg-cond}\sum_{l=1}^{N(\boldsymbol{f})}p_l(\boldsymbol{f})\int_{\gamma_{p_o,y_{p_o}-1}}^{\gamma_{p_o,y_{p_o}}}\cdot\cdot\cdot \int_{\gamma_{1,y_1-1}}^{\gamma_{1,y_1}}\mathrm{N}(\boldsymbol{W}^{(R)}\mid (\boldsymbol{D}(\boldsymbol{x},\boldsymbol{f})\boldsymbol{\beta}_{\pi_l(\boldsymbol{f})})^{(W)},\boldsymbol{\Sigma}^{(W)}_{\pi_l(\boldsymbol{f})})d\boldsymbol{W}^{(R)},\end{equation}
where the superscript $(W)$ indicates the portion of $\boldsymbol{D}(\boldsymbol{x},\boldsymbol{f})\boldsymbol{\beta}_{\pi_l(\boldsymbol{f})}$ and $\boldsymbol{\Sigma}_{\pi_l(\boldsymbol{f})}$ that corresponds to $\boldsymbol{W}^{(R)}$.

The weights in (\ref{eqn:cont-cond-X}) and (\ref{eqn:ord-reg-cond})
reveal additional flexibility that results from using the LDP.  Since the
weights are dependent on $\boldsymbol{\mathcal{F}}$, CMM-Mix is able
to capture relationships between $(\mathbf{Y}^{(R)},
\mathbf{Z}^{(R)})$ and $\boldsymbol{\mathcal{F}}$ beyond those implied
by $\boldsymbol{D}(\mathbf{X}^{(R)},
\boldsymbol{\mathcal{F}})$. This additional flexibility does not extend
to relationships between $(\mathbf{Y}^{(R)},
\mathbf{Z}^{(R)})$ and $\mathbf{X}^{(R)}$.

\subsection{Specifying the distance function $d(\cdot,\cdot)$ and the value of $d^*$}
\label{sec:distance}


We base $d(\cdot,\cdot)$ on Gower's generalized coefficient of
dissimilarity \citep{gower}, which is a standard dissimilarity measure
for mixed data \citep{kaufman,data-mining}.
The distance between two $q \times 1$ vectors $\boldsymbol{f}$ and $\boldsymbol{f}'$,
or $d(\boldsymbol{f},\boldsymbol{f}')$, is a weighted sum of
the element-wise distances, $\sum_{j=1}^q w_jd_j({f}_j,{f}_j')$.
Each $d_j$ takes values between $0$ and $1$, and
$\sum_{j=1}^qw_j=1$ with each $w_j\geq0$. For ordinal values,
$d_j({f}_j,{f}_j')=|{f}_j-{f}_j'|/(k_j^{(R)}-1)$. For continuous values, 
$d({f}_j,{f}_j')=|{f}_j-{f}_j'|/\mathrm{range}({f}_j)$. For nominal
values, we use the Hamming distance, $d_j({f}_j,{f}_j')=1$ when ${f}_j\neq
{f}_j'$, and $d_j({f}_j,{f}_j')=0$ when ${f}_j= {f}_j'$. Hence,
values of $d(\boldsymbol{f},\boldsymbol{f}')$ near zero indicate similar
$(\boldsymbol{f},\boldsymbol{f}')$, and values near one indicate
otherwise.  

When $\boldsymbol{\mathcal{F}}$ comprises survey design variables or
when $q$ is modest, 
we set
$w_j = 1/q$ for $j=1, \dots, q$.  This default assignment gives each
variable equal weight in determining similarities.  
However, one can improve computational efficiency and possibly inferential
accuracy by using only a subset of variables in 
$d(\boldsymbol{f},\boldsymbol{f}')$; that is, by setting $w_j=0$ for some $j$. 
In particular, it may be beneficial to set $w_j=0$ for 
variables in $\boldsymbol{\mathcal{F}}$ that do not contribute
meaningfully to predicting $\mathbf{X}^{(R)}$, since the primary
function of the LDP is to connect $\mathbf{X}^{(R)}$ and
$\boldsymbol{\mathcal{F}}$. Additionally, one can set 
$w_j=0$ for one or more $\boldsymbol{\mathcal{F}}$ variables
that are highly predictive of one another.

To perform feature selection \citep{guyon} and determine which $w_j=0$, we rely on values of
mutual information, which can be used to
describe dependencies between random variables of any type without
assumptions about the nature of their underlying relationships
\citep{battiti}.
For generic discrete random variables $A$ and $B$, 
their mutual information is $I(A,B)=\sum_{a}\sum_{b}
p(a,b)\log\{p(a,b)/(p(a)p(b))\}$; values near zero indicate only 
weak dependence between $A$ and $B$. 
In our context, the value of $I(\mathcal{F}_l,X_{j}^{(R)})$
for any $(l, j)$ is, intuitively speaking, the amount of
uncertainty in $X_j^{(R)}$ that is explained by $\mathcal{F}_l$. We
estimate these quantities to derive a single measure of the simultaneous explanatory power of 
$\mathcal{F}_l$ on $\mathbf{X}^{(R)}$, namely  
$I^{\mathrm{max}}_{\mathcal{F}_l,x}=\mathrm{max}\{I(\mathcal{F}_l,X_1^{(R)}),\dots,I(\mathcal{F}_l,X_{p_n}^{(R)})\}$.
Based on the estimated values of
$I^{\mathrm{max}}_{\mathcal{F}_l,x}$, we use  a forward selection 
procedure  to select the set of variables in $\mathcal{F}_l$ having
$w_l>0$, i.e., those deemed to have explanatory power beyond a threshold, and let the complementary set have $w_l=0$.  Details of the algorithm are in
Appendix \ref{sec:varselection}. 

We still include variables with $w_j=0$ in the regression for
$(\mathbf{W}^{(R)},\mathbf{Z}^{(R)})$.  If not, we would be forcing
these variables to be conditionally independent of the random
variables.  Even when mutual information values suggest weak
dependence with $\mathbf{X}^{(R)}$, ${\mathcal{F}}_j$ still may be
predictive of $(\mathbf{W}^{(R)},\mathbf{Z}^{(R)})$.  We note that
including irrelevant variables in
$\boldsymbol{D}(\mathbf{X}^{(R)},\boldsymbol{\mathcal{F}})$ is not 
problematic computationally (when $q$ is not huge), so that we prefer
not to force conditional independence {\em a priori}. 


We also must specify the neighborhood size $d^*$. As a benchmark,
consider what happens when $d^*\rightarrow 1^-$.  In this
case, all  $\boldsymbol{\Gamma}_h$ are in the
neighborhood of any value of $\boldsymbol{\mathcal{F}}$, so that $H_i$ in (\ref{eqn:hier_ldp}) is simply drawn from
$\{1,\dots,N\}$ as in the usual Bayesian mixture model. Hence, to
facilitate sharing of components by records with similar values of
$\boldsymbol{\mathcal{F}}$, we seek a $d^*$ away from one.
One approach is to determine a $d^*$ so
that each observation belongs to the same neighborhood as $r\%$ of the
observations on average, for instance $r=20$ \citep{chung}.  
Alternatively, we can base $d^*$ on  interpretations of the distance
function. For instance, with all nominal $\boldsymbol{\mathcal{F}}$,
setting $d^*=0.5$ implies that observations must exactly match on
at least $50\%$ of values to be in the same neighborhood and
share mixture components.  In our data applications, the results were
insensitive to different reasonable values of $d^*$.

\subsection{Posterior inference and missing data considerations}

For posterior inference, we use a Gibbs sampler based on the finite
stick-breaking representation of the DP \citep{ishjames}.
With missing values, the sampler proceeds via data augmentation, i.e.,
given a draw of the parameters, we draw new values of the missing
data. 
We present the  posterior full conditionals in  Appendix
\ref{sec:posterior}. After MCMC convergence, analysts can use the completed datasets for multiple imputation
inferences \citep{rubin, hu:mitra:reiter}, or directly make
posterior inferences from relevant functions of the parameters. 


Missing data  can cause numerical problems in the algorithm, namely that elements of
$\boldsymbol{\beta}_h$, $\boldsymbol{\Sigma}_h$, and
$\boldsymbol{\tau}$ can diverge towards very large values.  In
particular, problems occur when the mixture model samples clusters
with all observed values of some ordinal variable equal to the first
or last category, e.g., $Y_{ij}^{(R)}=1$ for all observed
$Y_{ij}^{(R)}$ in some cluster.  This creates a perfect prediction
problem and ensuing estimation difficulties. In our example, the
mixture model will strongly favor imputing category 1 for the missing values of $Y_{ij}^{(R)}$ in the
cluster. Hence, it will favor making the corresponding latent $W_{ij}$
very negative, which happens when elements of $\boldsymbol{\beta}_h$
are large in magnitude. This in turn 
can cause some elements of $\boldsymbol{\tau}$ to become extremely
large.  Related problems arise with other types of perfect predictions
within clusters. We note
that these problems can arise in any mixture model (not just
CMM-Mix) that uses probit specifications for ordinal variables. 

We use a quick and dirty fix that prevents elements of
$\boldsymbol{\beta}_h$ from getting too large. Given the standardization of all continuous variables and
choice of cut-offs for the latent continuous random variables,
practically we need not allow elements of $|\boldsymbol{\beta}_h|$ to
exceed 4 or 5, as this still allows the average value of each $\mathbf{Y}_j^{(R)}$ to be
as small as $1$ or as large as $k_j^{(R)}$. We therefore restrict
${\tau}^2_j \leq 6$ via a truncated inverse-gamma prior in place of an
inverse-gamma prior. This keeps the parameters from diverging to large
values without being overly restrictive, since the
draws for $\tau_j^2$ are centered far to the left of the truncation
value in all of our model implementations. 
We include plots of posterior samples for $\tau_j^2$ in the supplementary
material. 

\section{Evaluating CMM-Mix using a data fusion scenario} 
\label{sec:DataFusion}

In data fusion, analysts seek
to combine information from two or more databases containing
information on disjoint sets of individuals.
For example, a set of demographic
variables $A$ is available in two databases $D_1$ and $D_2$, a set of variables
$B_1$ is available only in $D_1$, and a set of
variables $B_2$ is available only in $D_2$.  The analyst seeks to use $D_1$ and $D_2$ to 
learn about the joint distribution of all variables.  Without
simultaneous observations of $\{A,B_1,B_2\}$, the analyst is forced
to make identifying assumptions about the conditional
associations between $B_1$ and $B_2$ given $A$. The simplest and most
common assumption is that $B_1$ and $B_2$ are conditionally
independent given $A$, which may be reasonable when $A$ is rich.

\citet{Kamakura} proposed that data fusion be implemented by fitting mixture
models to the concatenation of $(D_1, D_2)$, so as to capture
nonstandard distributions and complex associations among $(A, B_1)$
and $(A, B_2)$ automatically.
However, \citet{gilula} questioned whether or not mixture models for data fusion truly
encode conditional independence between $B_1$ and $B_2$.  Indeed, 
\citet{fosdick} find in simulations that a mixture model 
generated stronger 
estimated associations between $B_1$ and $B_2$ than implied by conditional independence. 

These results motivate the simulation studies of this section.  Specifically,
we investigate whether or not  CMM-Mix more
faithfully respects conditional independence than a fully joint
mixture model in data fusion contexts. Viewed more generally, we compare how well CMM-Mix and a
fully joint mixture model estimate true joint distributions in the presence of
missing data.  We also performed a separate
empirical study with arbitrary itemwise missing data patterns; results
and conclusions, available in the supplementary material, indicate that CMM-Mix provides accurate inferences. 


\subsection{Constructing the data fusion scenario}


We base the simulations on data from a survey of
$n=3567$ individuals collected by the book publisher HarperCollins.
We treat eleven variables from the survey as $A$ variables, including six ordinal
variables (age, passion for books, reading hours, income, passion for
Internet, and opinion on romance in books) and five nominal variables
(work status,  importance that books that challenge the reader, laptop
ownership, eBook reader ownership, use of audiobooks).  To construct a data fusion scenario
where conditional independence is known to hold, we generate a continuous variable $Z$, an ordinal variable $Y$, and 
a nominal variable $X$ such that $f(Z,Y,X \mid A) = f(Z \mid A) f(Y \mid A) f(X \mid A)$. We use normal, probit, and multinomial logistic 
regressions to generate $Z$, $Y$, and $X$, respectively, using various main effects and interactions involving only $A$ in the predictor functions.
The exact specifications are presented in the supplementary material.  
After binding these simulated variables to $A$, we blank $(X, Z)$ for the first 1189 rows, $(X, Y)$ for the second 1189 rows, and $(Y,Z)$ for the 
final 1189 rows.  This simulates a data fusion scenario with three databases. This process is repeated 40 times to create 40 unique sets of databases to be fused, 
which results in Monte Carlo standard errors that are adequately small for our purposes. 




\subsection{Results}

We consider the eleven $A$ variables as $\boldsymbol{\mathcal{F}}$,
and $(Z, Y, X)$ as $\boldsymbol{\mathcal{R}}$. 
In one randomly chosen set of simulated $\boldsymbol{\mathcal{R}}$, the (normalized) values of $I^{\mathrm{max}}_{\mathcal{F}_l,x}$ to two decimal places 
in descending order are $(.25, .23, .07, .07, .06, .06, .02, .02, .01, .01,.00)$; other replications yield similar values. Given the number of weak associations, 
 we investigate two possibilities for assigning non-zero weights in $d(\cdot, \cdot)$.  First, we allow  
only the top two scoring variables---reading hours (ordinal with 5 categories) and laptop ownership (binary)---to have $w_j>0$; this includes only moderately strong predictors.  Second, we allow the top six
scoring variables---ordinal variables age, passion for books, and income, and binary variable
 desire for challenge in reading---to have $w_j>0$; this  includes variables that are weakly predictive.
The maximum normalized mutual information between any $(\mathcal{F}_l,\mathcal{F}_{j})$ pair is $0.27$, corresponding to passion for books and reading hours per week, suggesting 
no serious redundancies among $\boldsymbol{\mathcal{F}}$.



For each feature selection, we consider three sets of possible $d^*$ values.  For the two feature model, we consider $d^* \in (0.5, 0.25, 0.125)$. 
Here, $0.5$ is the maximum distance between two individuals' vectors of reading hours and laptop ownership when only one of the two variables matches.
The distance is $0.125$ when the laptop ownership matches and reading hours is separated by just one category. This
 is the most stringent criterion besides requiring that both variables must match exactly, which would set $d^*=0$. The $0.25$ is
an intermediate value.
These $d^*$ values imply respectively that each observation is in the same neighborhood as $66\%$, $48\%$ and $34\%$ of the other observations
 on average.
For the six variable feature selection, we consider $d^* \in (0.375, 0.30, 0.25)$.  These $d^*$ values imply that each
 observation is in the same neighborhood as $52\%$, $34\%$, and $23\%$ of the other observations on average.


We fit the six versions of CMM-Mix to each concatenated dataset on $\{Y,Z,X,A\}$ containing missing values.  We refer to the two feature models as C-2S, C-2M, and C-2L, where 
the S, M, and L indicate the smallest to largest values of $d^*$, respectively. We refer to the six feature models as C-6S, C-6M, and C-6L 
using analogous nomenclature. We also fit a fully joint mixture model that puts all fourteen variables in $\boldsymbol{\mathcal{R}}$.
For each method, we create $10$ completed data sets for use in multiple imputation inferences.   

We begin with inferences for all 
285 cell probabilities associated with the bivariate distributions of $(X,A_j)$ and $(Y,A_j)$ for all $j$. 
For each of the 40 simulated datasets, we calculate the proportion of multiple imputation $95\%$ confidence intervals that contain their corresponding ground truths. We also calculate summaries of the absolute errors of the point estimates, including the mean, $25$th percentile, and $75$th percentile.
As evident in Table \ref{table:coverage_AC},
the joint model results in the lowest coverage rates and largest absolute errors.  Differences are unlikely due to 
Monte Carlo error, as the standard errors of reported quantities across the 40 sets of results are relatively small; for example, 
the estimated standard errors of all coverage rates for the conditional models are less than .006, and the standard error of the  
coverage rates for the joint model is .016. 

The models with small $d^*$ values result in the 
largest coverage rates. The absolute errors are generally smallest
under the small and medium $d^*$ values,  but all CMM-Mix models
tend to be more accurate than the joint model. The models with small
$d^*$ outperform those with large $d^*$, highlighting the  
 benefits of dependent cluster assignments.   

\begin{table}[t]
\centering
\begin{tabular}{l|c|c|c|c|c|c|c}
 & Joint & C-2S & C-2M  & C-2L & C-6S & C-6M & C-6L  \\ \hline

 \% of CIs Covering & .655 & .905 & .813 & .792 & .854 & .839 & .791 \\
 Mean Abs. Error & .0128& .0074 & .0066  & .0081 & .0066 & .0074  & .0087 \\
$25\%$ile Abs. Error & .0025 & .0016 & .0012  & .0011 & .0013 & .0013  & .0013  \\
$75\%$ile Abs. Error & .0161 & .0097 & .0086 & .0090 & .0089 & .0091 & .0100
\end{tabular}
\caption{Summary of inference for cells corresponding to bivariate
  discrete distributions with missing data. Entries include the
  proportions of $95\%$ multiple imputation confidence intervals (CIs) that
  contain the truth, and the mean, $25$th percentile, and $75$th percentile of absolute errors, averaged over all 40 datasets.}
\label{table:coverage_AC}
\end{table}

Turning to $Z$, inferences for its mean, $25$th percentile, and $75$th
percentile are more accurate under the CMM-Mix models 
than the joint model.
Almost all $95\%$ CIs from the conditional models contain the true mean zero, 
with all average point estimates less than $.027$ in absolute value. By comparison, the joint model produces an average point estimate of $.093$, 
and more than half of the $95\%$ CIs do not contain zero. 
We also 
estimate $\mathrm{E}(Z \mid A_j)$ for each value of $A_j$ for all $j$.
Once again, the joint model results in noticeably larger errors 
than the CMM-Mix models. The average absolute errors of the 46 point estimates from the joint model
versus C-2S are shown in Figure \ref{fig:Acont_cond_C_2S}. Figures for 
other models are in the supplementary material. 


\begin{figure}[t]
\centering
\includegraphics[scale=0.5]{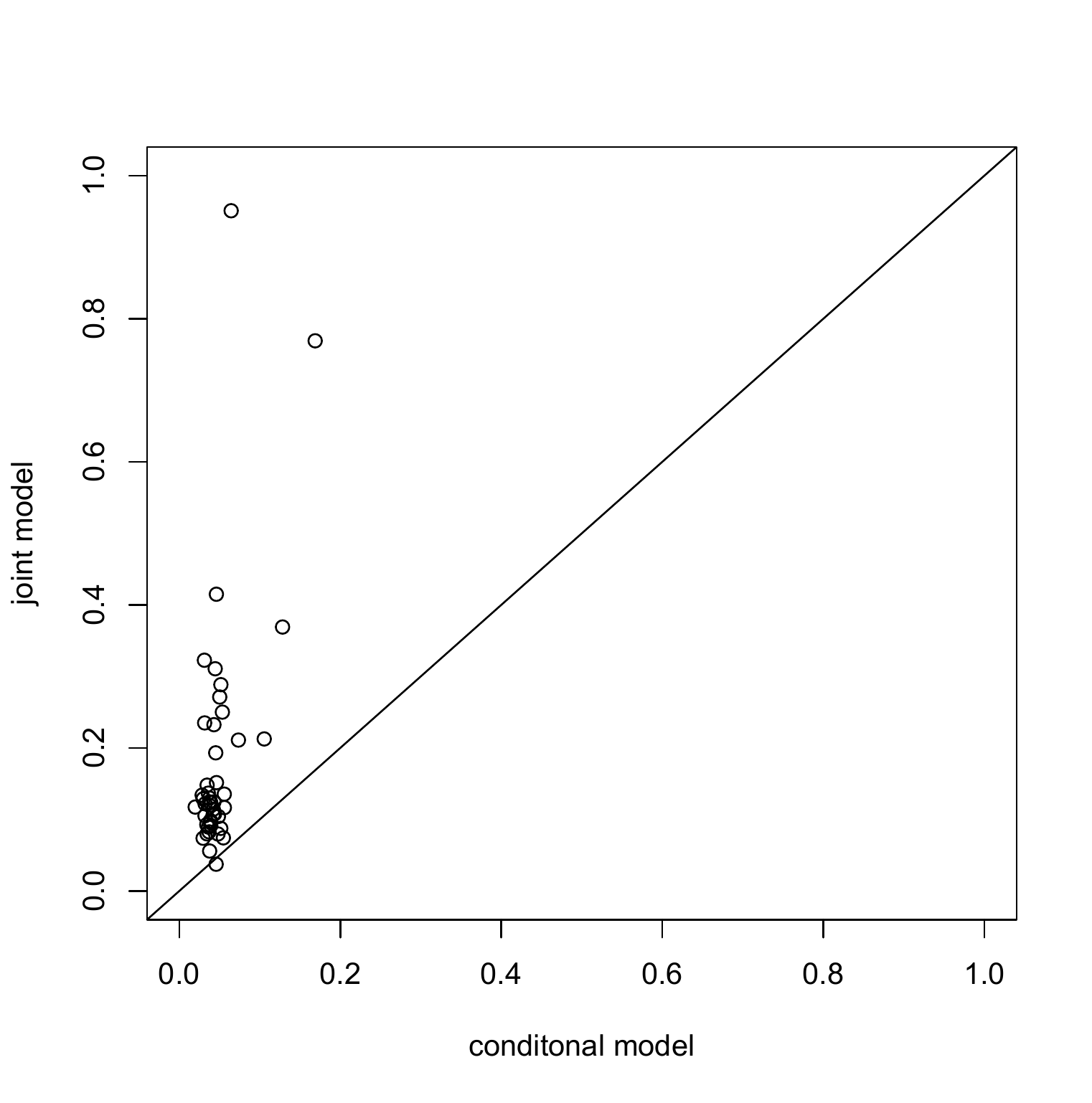} 
\caption{Average absolute errors of 46 multiple imputation point estimates for
  the mean of $Z$ conditional on each $A$ variable from the joint
  model and C-2S.}
\label{fig:Acont_cond_C_2S}
\end{figure}

We next investigate which mixture models best estimate the 
generation model for $(X, Y, Z) \mid A$; in other words, which
models are 
most faithful to the conditional independence assumption for data fusion. Here we focus on one randomly selected simulation run for simplicity in presentation of results. For each of the 10 completed data sets, we fit a regression model of $Z$ conditional 
on $X$ and $Y$, as well as all variables in $A$ that actually generated the data.
We use multiple imputation inference to create point estimates and $95\%$ confidence intervals for all regression
 coefficients. We also fit the model to the fully observed data without any missing values, referred to as the pre-missing data, 
and obtain the true empirical $95\%$ confidence intervals for regression coefficients. 

With the full joint model, the 95\% confidence intervals for three of
the four regression coefficients for $X$ and $Y$ do not contain the
true value of zero, whereas with  
all versions of CMM-Mix and the pre-missing data all intervals contain zero.  All models result in confidence intervals that exclude 
zero for ten of the eleven non-zero coefficients involving $A$. 
The average absolute error of the regression coefficient estimates
from the joint model is $.252$, whereas the averages are less than $.08$
for all CMM-Mix models. 
Eleven of the sixteen regression coefficient confidence intervals from the joint model do not even overlap with their
corresponding confidence intervals 
based on the pre-missing data; in contrast, all sixteen confidence intervals from CMM-Mix fully contain their corresponding 
confidence intervals based on the pre-missing data.
We find similar results when fitting regression models that treat $Y$ or $X$ as the response.
We also validate that the joint model is less consistent with conditional independence assumptions than the CMM-Mix models 
using the strategy of \citet{Kunihama} based  on conditional mutual information.  Details of all additional results are in the 
supplementary material.

Finally, we investigate the performance of statistical matching \citep{putten2002,wicken2008}, a common technique used for data fusion. 
Here, we create fused data sets via exact matching on the 11
$A$ variables using the \texttt{StatMatch} package in R
\citep{StatMatch}.
For each record $i$ with missing $Y_i$, this method identifies all
observations $\{i' \neq i: Y_{i'} \textrm{ is observed}\}$ with the
smallest Hamming distance from $A_i$, and samples one of their $Y$
values as an imputation for the missing $Y_i$.  It operates similarly
for observations missing $X$ or $Z$.
Using this method, on average, only $65\%$ of confidence intervals associated with bivariate probabilities contain the true values.
Thus, the statistical matching approach is not as effective as the conditional mixture model approaches.

In summary, the results suggest that the imputations for the variables with missing data  
from the joint mixture model are not consistent with conditional
independence, whereas those from the CMM-Mix models are.  Additionally, the CMM-Mix models estimate  
the relationships in the data-generating model as or more reliably than the joint model.
There is  evidence of potential for inferential gains by choosing smaller $d^*$
 and a more parsimonious set of features driving the cluster
 assignments.

\section{Conditional Inference from a Quota Sample}
\label{sec:HarperCollins}

We now analyze a quota sample conducted by HarperCollins in 2013 to learn about U.\ S.\ consumers'
reading behaviors and interests. The sample comprises $n=3631$ individuals, sampled to achieve fixed quotas in various  
age, gender, ethnicity, and region (location) groups.  We seek inferences for the 
relationships among income, passion for books, passion for the Internet, reading hours, laptop ownership,
 eBook ownership, and desire to be challenged in reading. Most variables are complete; only income and reading hours have 
isolated missingness, which we assume to be at random.  Income, passion for books and the Internet, and reading hours are
ordinal and comprise $\mathbf{{Y}}^{(R)}$;  eBook ownership and laptop ownership are nominal and comprise $\mathbf{{X}}^{(R)}$; and,
the measure of desire for challenging books is numerical and comprises $Z^{(R)}$.  We condition on all the design variables when 
fitting CMM-Mix, letting $\boldsymbol{\mathcal{F}}$ include age,
gender, ethnicity and region with all $w_j>0$.

We fit CMM-Mix with $d^*=0.25$, which implies that each observation is in the same neighborhood as $14\%$ of the 
other observations on average, and observations must match on least of 3 of the 4 design variables to be in the 
same cluster. Posterior predictive model checks do not suggest
evidence of lack of model fit; results are in the supplementary material.  
We focus on full posterior inference rather than multiple imputation inference, as some
sub-groups (i.e., combinations of age, gender, ethnicity, region) have few or no individuals in sample. 
We make conditional inferences rather than marginal inferences, as we
do not have population distributions of the quota variables.

We focus inferences on individuals aged 25--34, one of the most sought after demographics in 
marketing \citep{atlantic,business-insider}. 
In particular, we contrast the characteristics of individuals in this group who own eBook
 readers and those who do not own eBook readers.  CMM-Mix enables inferences about many other demographic groups; we 
report only on one here to present a concise and coherent analysis.

\begin{figure}
\centering
\begin{tabular}{cc}
\includegraphics[height=3.0in,width=3in]{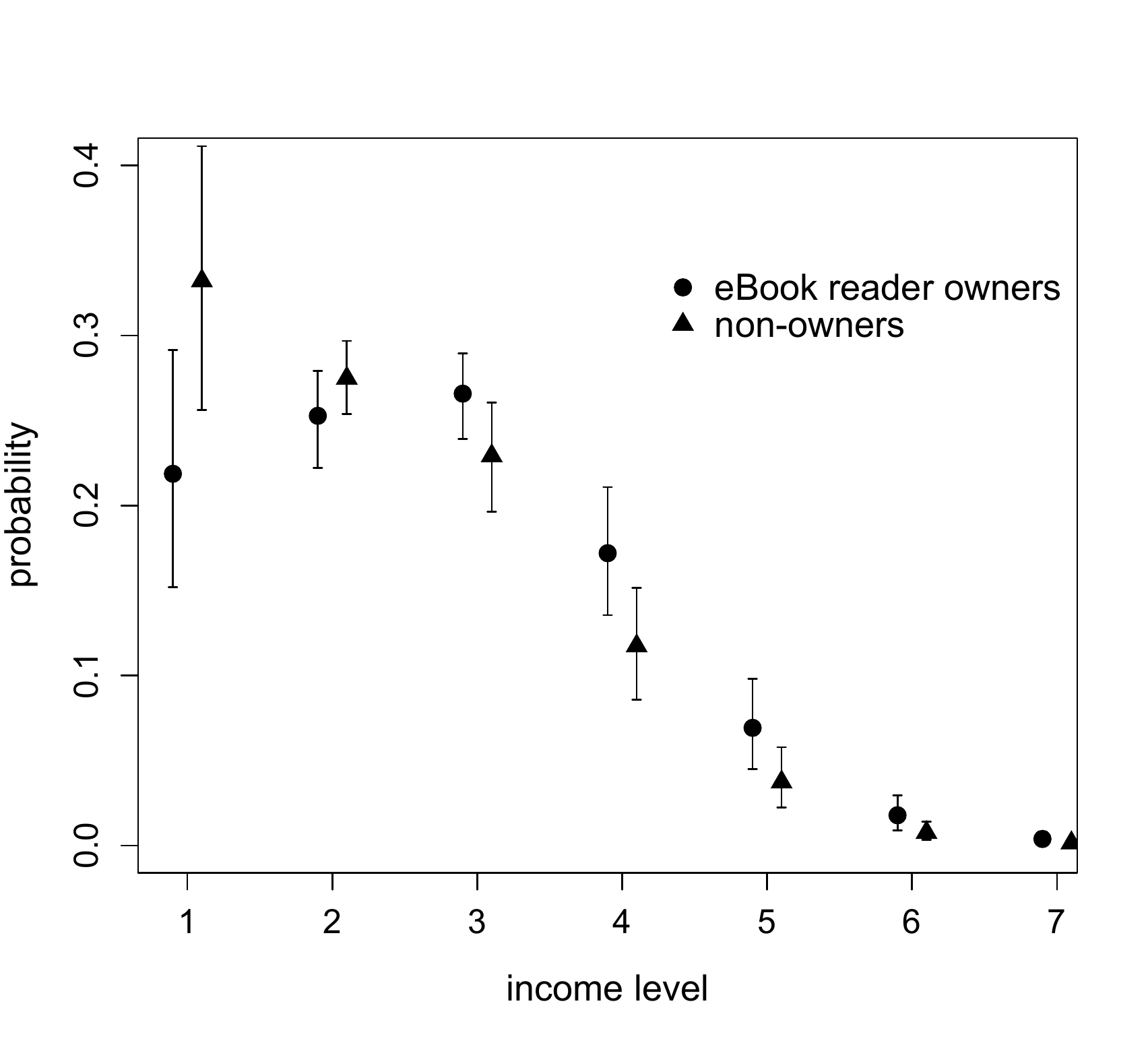}&
\includegraphics[height=3.0in,width=3in]{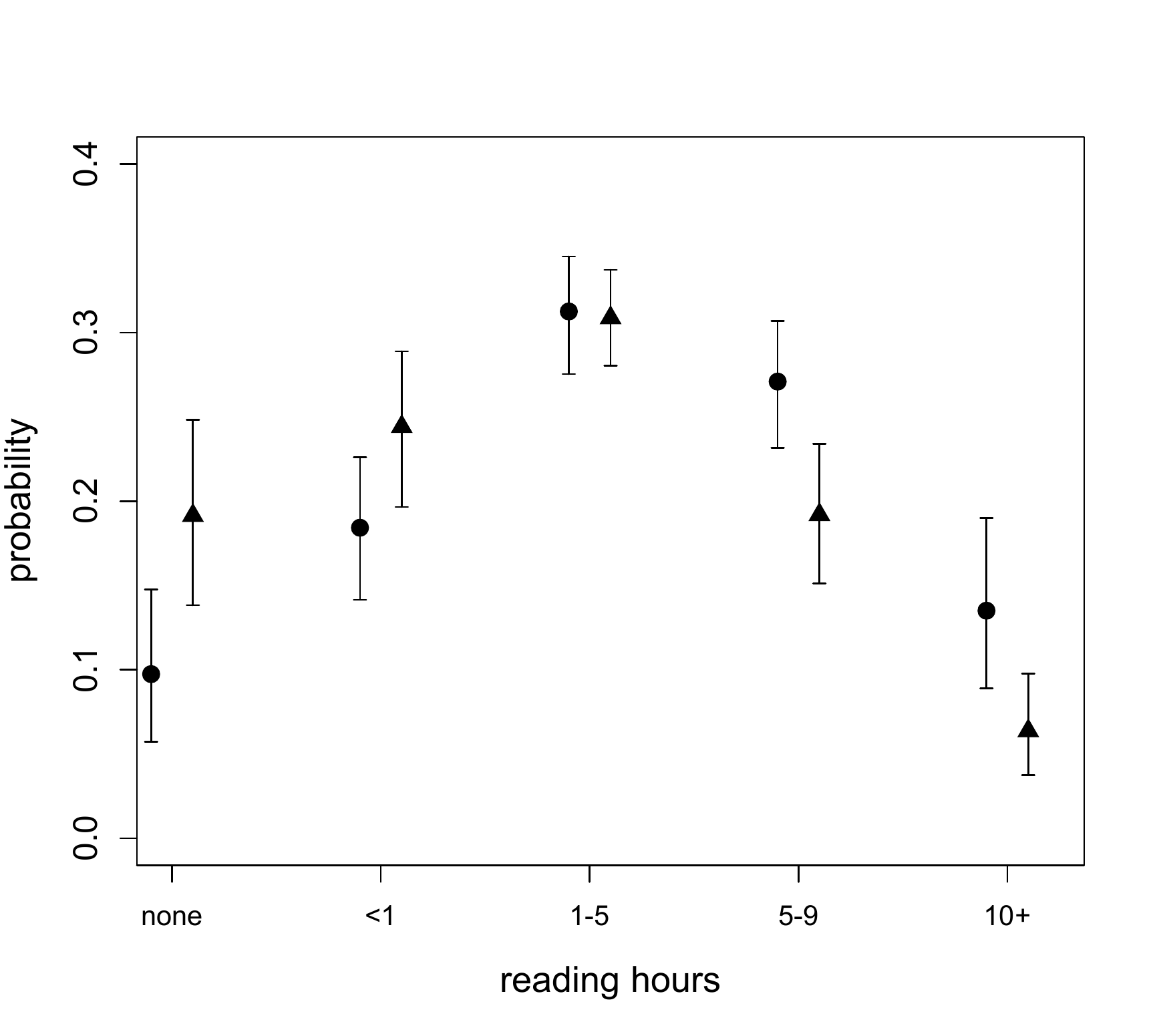}
\end{tabular}
\caption{Left: Posterior mean and $90\%$ credible intervals for the
  probability assigned to each income level for eBook owners (circle
  symbols) and non-owners (triangle symbols). Income levels 1 through
  7 correspond to $<25$K, $25-44$K, $45-74$K, $75-99$K, $100-149$K,
  $150-199$K, $\geq200$K. Right: Posterior mean and $90\%$ credible
  intervals for the probability assigned to each level of weekly
  reading hours for owners versus non-owners. Both displays refer to white females aged 25--34 living in the South.}
\label{fig:inc-cond-eBook}
\end{figure}

Figure \ref{fig:inc-cond-eBook}  displays the distribution of income for eBook reader owners and non-owners 
for white females aged 25--34 living in the South.  Owners of eBook
readers tend to have higher incomes than non-owners, with the most
striking difference in the percentage of individuals making less than \$25000 per year.   
We obtain these posterior inferences using expression (\ref{eqn:ord-reg-cond}), 
integrating out the nominal random variable laptop, ${X}^{(R)}_2$, using the estimates for 
$\mathrm{Pr}({X}_2^{(R)}={x})=\sum_{l=1}^{N(\boldsymbol{\mathcal{F}})}p_l(\boldsymbol{\mathcal{F}})\psi^{(2)}_{\pi_l(\boldsymbol{\mathcal{F}}),x}$.
Figure \ref{fig:inc-cond-eBook} also displays the posterior
distribution of weekly reading hours for owners and
non-owners. Overall, owners  of eBook readers tend to spend more hours
reading than non-owners, and are noticeably less likely to spend zero
hours reading. 


\begin{figure}
\centering
\includegraphics[height=6in,width=6in]{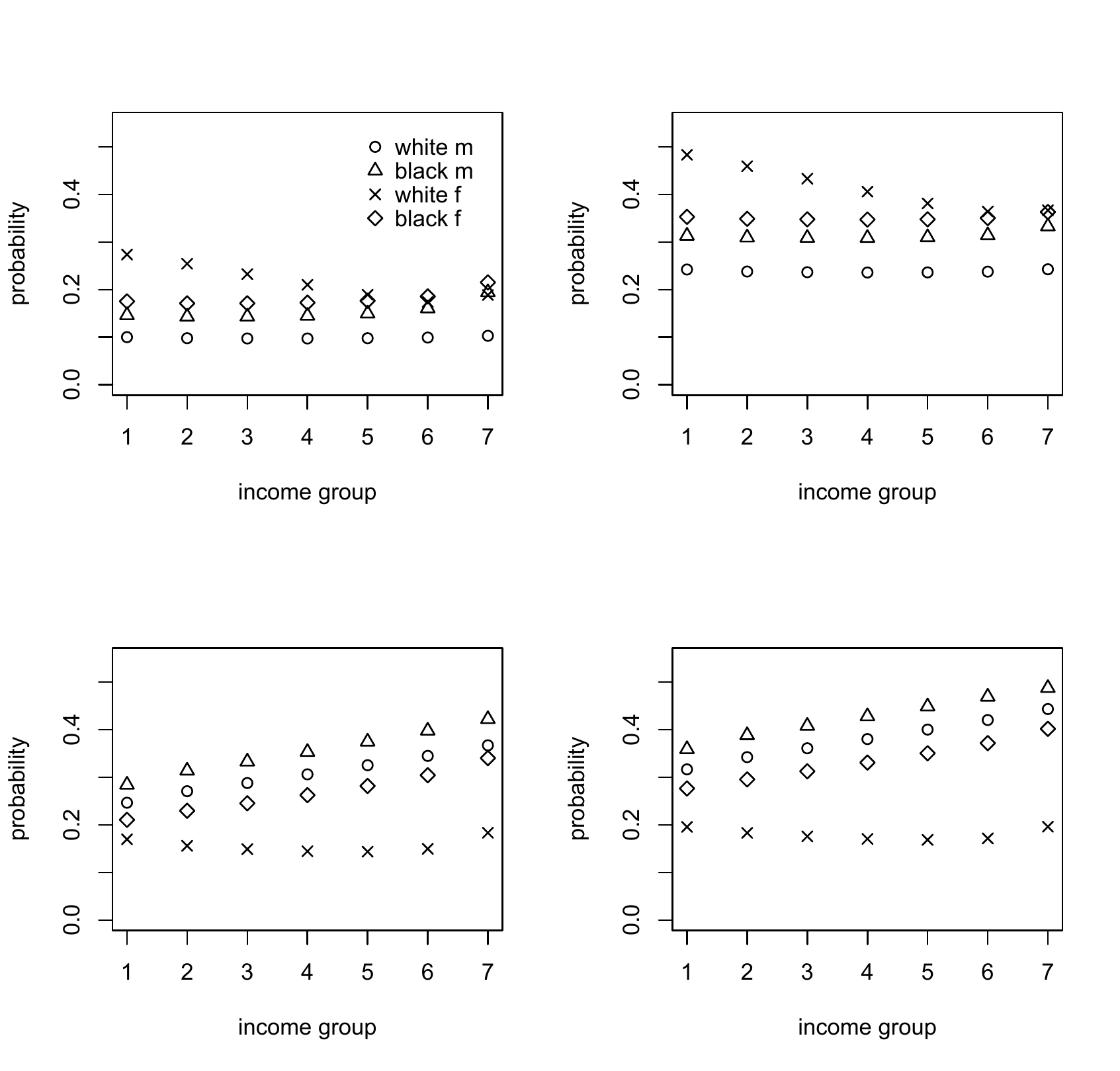}
\caption{Posterior mean estimates for the probability that one views books as very important (top row) and for the probability that one views the Internet as very important (bottom row) as a function of income, gender, and race, for eBook non-owners (left column) and owners (right column).
Income levels 1 through 7 correspond to $<25$K, $25-44$K, $45-74$K,
$75-99$K, $100-149$K, $150-199$K, $\geq200$K. }
\label{fig:passion-cond-eBook-income}
\end{figure}

These results are in accord with prior research suggesting that higher
income individuals are more likely to be frequent readers
\citep{pew,pew2}. Research also suggests that higher income
individuals use the Internet more frequently.
Building on these findings, we next contrast eBook owners' and
non-owners' views on the importance of books and of the Internet,
drilling down by race to provide additional information.
Figure \ref{fig:passion-cond-eBook-income} displays the probability
that individuals aged 25--34 from the South regard books as very
important in their lives, and the probability that
individuals aged 25--34 regard the Internet as very important in their
lives, as functions of gender, race, and  eBook reader ownership.
The posterior inferences suggest that, regardless of income level, white females are most likely to
view books as important, and white males are the least likely to view
books as important.  As might be expected, those who own eBook readers
are more likely to view books as very important than those who do not
own eBook readers.  The posterior inferences also suggest that,
regardless of income level, black males are most likely to view the Internet
as important, and white females are least likely to view the Internet
as important. The results suggest that the importance of the Internet
tends to increase as a function of income for both eBook reader owners
and non-owners, except for white females. The model suggests 
little to no interactions between gender and eBook
ownership, and race and eBook ownership. Exploratory analysis and posterior
predictive checks are consistent with these findings, indicating this is not some artifact
of the model specification.


Finally, we examine whether or not eBook reader owners prefer books that
challenge them to think more than those who do not own eBook readers.
This involves posterior distributions of the continuous
variable ``challenge'', conditional on eBook reader ownership for each combination of age, gender, ethnicity, and region.  
Larger values of ``challenge'' indicate increasing appreciation for
books that challenge the reader to think. The distribution $f(Z^{(R)}
\mid \boldsymbol{\mathcal{F}}, \mathbf{X}^{(R)}) =  
\sum_{l=1}^{N(\boldsymbol{\mathcal{F}})}p_l(\boldsymbol{\mathcal{F}})\mathrm{N}(Z^{(R)}\mid
(\boldsymbol{D}(\mathbf{X}^{(R)},\boldsymbol{\mathcal{F}})\boldsymbol{\beta}_{\pi_l(\boldsymbol{\mathcal{F}})})^{(Z)},\boldsymbol{\Sigma}^{(Z)}_{\pi_l(\boldsymbol{\mathcal{F}})})$, 
where the superscript $(Z)$ indicates the portion of
$\boldsymbol{D}(\mathbf{X}^{(R)},\boldsymbol{\mathcal{F}})\boldsymbol{\beta}_{\pi_l(\boldsymbol{\mathcal{F}})}$
and $\boldsymbol{\Sigma}_{\pi_l(\boldsymbol{\mathcal{F}})}$ that
corresponds to $Z^{(R)}$. To estimate $f(Z^{(R)}\mid
X^{(R)}_1,\boldsymbol{\mathcal{F}})$, where $X^{(R)}_1$ refers to the
random variable ``eBook ownership'', we  integrate out ${X}^{(R)}_2$ using the estimates for
$\mathrm{Pr}({X}_2^{(R)}={x})=\sum_{l=1}^{N(\boldsymbol{\mathcal{F}})}p_l(\boldsymbol{\mathcal{F}})\psi^{(2)}_{\pi_l(\boldsymbol{\mathcal{F}}),x}$. 

\begin{figure}
\centering

\includegraphics[height=2.0in,width=6.2in]{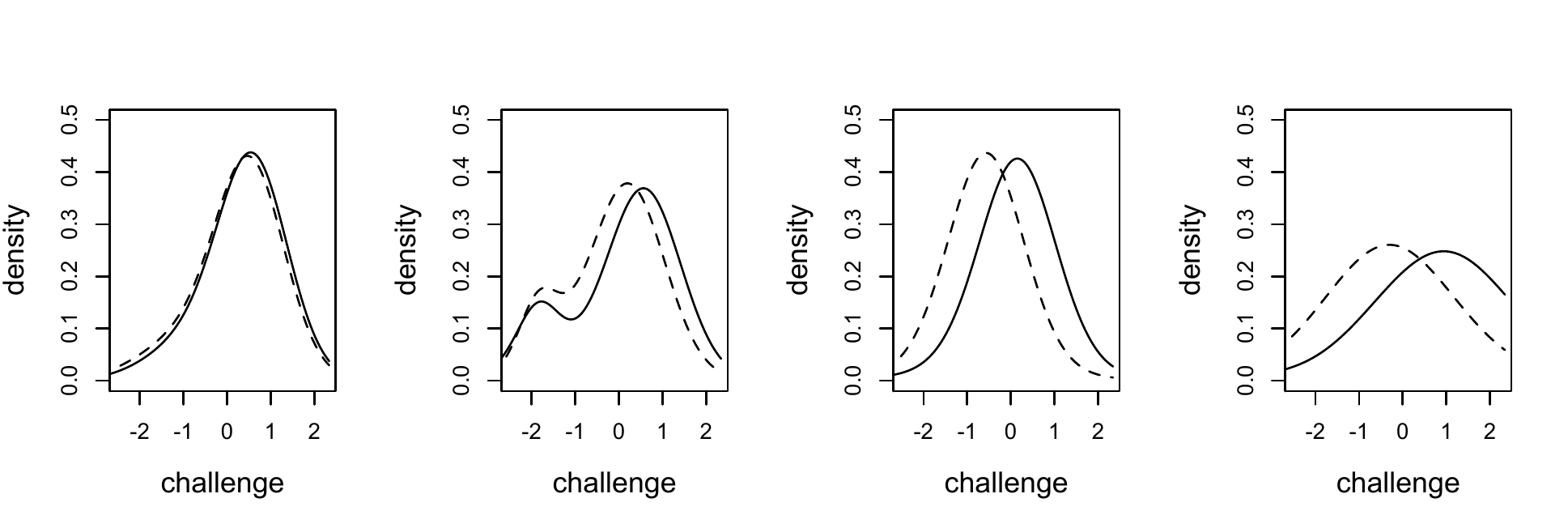}

\caption{Mean posterior distribution of the ``challenge''
  variable given that one owns an eBook (solid) or does not own an eBook (dashed). Figures refer to four combinations of age, gender, ethnicity, and region. Left to right: 25--34, female, white, South; 25--34, male, white, Northeast; 35--44, female, Hispanic, West; 45--54, male, other, West.
} 
\label{fig:challenge}
\end{figure}

Figure \ref{fig:challenge} displays mean posterior 
distributions of the challenge variable for four combinations of age, gender, ethnicity, and
region. The shapes of the distributions are varied, with some having
skewness or bimodality and others being standard unimodal
distributions. For almost all combinations of age, gender, ethnicity,
and region, the distribution of the challenge variable is centered on
larger values for eBook reader owners than for non-owners. This
suggests that those who own eBook readers tend to prefer books that
are challenging as compared to those who do not own eBook
readers. 
 
All of these inferences are conditional on fixed quota variables. For
unconditional inferences, for instance national estimates, analysts
can integrate out $\boldsymbol{\mathcal{F}}$ over its distribution in
the population. 
In our context, this can be accomplished by estimating
the quantity of interest for each combination of age, gender,
ethnicity, and region, and averaging over these estimates according to
each group's population share.
For inferences conditional on only age and region, for example, we  marginalize over the distribution of
ethnicity and gender in the population of U.\ S.\ individuals aged 18
and older.  


\section{Discussion}
\label{sec:Discussion}

We conclude with a discussion of future research directions, beginning
with the feature selection algorithm for the LDP.  
The simulations suggest that when the number of variables in $\boldsymbol{\mathcal{F}}$ is large,
one can improve accuracy by selecting
only a subset of the variables to include in the distance
calculation. We used equal weights for features selected to be in the distance function. It may be beneficial to 
allow the non-zero weights to differ, for example to be proportional
to the values of mutual information with the $\mathbf{X}^{(R)}$
variables. 

The CMM-Mix models also may point an alternative path for handling
survey weights in Bayesian mixture models inference, which is an active area of
research \citep[e.g.,][]{kunihama:herring,si:pillai,savitsky:toth}. Since the survey
weights are fixed and fully observed, one could include them as an
$\mathcal{F}$ variable. This would make the conditional
distribution for the random variables a function of the survey weights, as is
done in the Gaussian process model of \citet{si:pillai}.  Here,
observations with similar survey weights would be encouraged  to share
mixture components. 

Finally, our work with CMM-Mix reveals problems when using mixture
models to impute missing ordinal values that can be explored further. 
Specifically, the mixture models sometimes favor clusters that are
homogeneous in observed values, which encourages imputations also to be
at the observed values.
This can be undesirable, as it may result in underestimation of
imputation uncertainty, particularly when cluster assignments are
relatively stable.  
It may be beneficial somehow to add uncertainty into
homogeneous clusters. Related ideas are used by \cite{paiva}, who
adjust the weights associated with mixture components in order to
generate imputations that follow a nonignorable missing data mechanism.

\bibliographystyle{asa}
\bibliography{biblioC}

\appendix
\section{Appendix}
\subsection{Prior Specification}\label{sec:prior}

The particular matrix-normal base distribution for the regression coefficient matrices $\boldsymbol{\beta}_h$ implies that elements of $\boldsymbol{\beta}_h$ in the same row have the same variance, but the variance differs for elements in different columns. This is a reasonable assumption because the variables in $\mathbf{Z}^{(R)}$ are standardized, and the cut-offs $\gamma_{j,1}$ and $\gamma_{j,k_j^{(R)}-1}$ for $j=1,\dots,p_o$ can be fixed to values that imply all variables in $(\mathbf{W}^{(R)},\mathbf{Z}^{(R)})$ have the same scale. However, we want to allow the variances of regression coefficients to differ across covariates, hence the use of the $\boldsymbol{\tau}=(\tau_1^2,\dots,\tau_k^2)$ vector. 

We assume a conjugate gamma distribution for the precision parameter
$\alpha$. The number of effective global components in the mixture is
influenced by $\alpha$.  Thus, the shape and rate parameters can be fixed to small values, such as $0.5$, to represent a relatively noninformative prior on the number of global mixture components. We recommend fixing $a_1=\dots=a_{d_j}=1$ in the Dirichlet base distribution so that the probability vectors $\boldsymbol{\psi}_h^{(j)}$ can encompass a wide variety of values.

To specify priors for the remaining parameters, consider the limiting case of the mixture 
as $\alpha\rightarrow 0^+$, which results in a single multivariate normal distribution 
for $(\mathbf{W}^{(R)},\mathbf{Z}^{(R)})$. Our objective is to center and scale the mixture kernel appropriately. Since the $\mathbf{Z}^{(R)}$ are standardized, fixing $\gamma_{j,1}\approx-3$ and $\gamma_{j,k_j-1}\approx 3$ places each $\mathbf{Z}^{(R)}_j$ and $\mathbf{W}^{(R)}_j$ on similar scales.
 We use a conjugate matrix-normal prior for $\boldsymbol{\beta}_0$,
 assuming
 $\boldsymbol{\beta}_0\sim\mathrm{MN}(\boldsymbol{0},\boldsymbol{I},h\boldsymbol{I})$. This
 represents prior information that the regression coefficients are
 centered at zero. We assume $\tau_j^2\sim
 \mathrm{IG}(a_{\tau},b_{\tau})$, possibly truncated to the lie below
 a finite value (e.g., 6) if numerical problems arise, and $\boldsymbol{S}\sim \mathrm{W}(a_S,\boldsymbol{B}_S)$.  Let $v_j$ denote an estimate for the variance of $(\boldsymbol{W}^{(R)},\boldsymbol{Z}^{(R)})_j$, which is given by $((\gamma_{j,k_{j}^{(R)}-1}-\gamma_{j,1})/4)^2$, for ordinal variables $j=1,\dots,p_o$. For continuous variables $j=p_o+1,\dots,p_o+p_c$, the quantity $(\mathrm{range}(Z^{(R)}_{j-p_o})/4)^2$ can be used as a proxy for variance. Under the standardization and cut-off points described, each $v_j\approx 1.5^2$.
The marginal prior variance for $(\boldsymbol{W}^{(R)},\boldsymbol{Z}^{(R)})_j \mid \boldsymbol{D}$ is
$a_S(\boldsymbol{B}_{S})_{jj}(\nu-p_c-p_o-1)^{-1}+(b_{\tau}(a_{\tau}-1)^{-1}+h)\sum_{l=1}^kD_l^2$.
For a default prior, we fix $a_\tau$, $a_S$, and $\nu$ to relatively
small values. Because the design vector $\boldsymbol{D}$ consists of a
one for the intercept term followed by only zeroes and ones, the
smallest value it can take is $1$. To be conservative, we fix
$\sum_{l=1}^KD_l^2$ to $1$, and determine values of
$\boldsymbol{B}_S$, $b_{\tau}$, and 
$h$ such that marginal prior variance is approximately equal to
$v$. For instance, we can set $\boldsymbol{B}_s=(\nu-p_c-p_o-1)(3a_{S})^{-1}\mathrm{diag}(v,\dots,v)$, $b_{\tau}=3^{-1}(a_{\tau}-1)v$, and $h=3^{-1}v$. This implies a prior variance for each element of $\boldsymbol{\beta}_h$ that allows a typical value of $Y_{j}^{(R)}$ to be as small as $1$ or as large as $k_j^{(R)}$. It also centers each diagonal element of $\boldsymbol{\Sigma}_h$ at $0.75$, which is reasonable given that each $W_j^{(R)}$ and $Z_j^{(R)}$ range from around $-3$ to $3$. Although this can be used as a default approach to prior
specification, in our experience inferences like density
estimates and regression functionals are insensitive to the choice of
values for the parameters of these hyperpriors.

\subsection{Algorithm for Selecting Variables in $d(\cdot, \cdot)$}
\label{sec:varselection}

We select variables in $\boldsymbol{\mathcal{F}}$ in accordance with the
mutual information of each $\mathcal{F}_l$, where $l=1,\dots,q$, with
$\mathbf{X}^{(R)}$, as well as the mutual information between pairs of
$(\mathcal{F}_l,\mathcal{F}_{l'})$, $l\neq l'$. \citet{ding:peng} and
\citet{peng} provide an algorithm referred to as
minimal-redundancy-maximal-relevance (mRMR) which we adapt for our
purposes, extending their algorithm to handle
multivariate $\mathbf{X}^{(R)}$.  The algorithm proceeds as follows.
\begin{enumerate}
\item Set
  $\boldsymbol{S}=\{\emptyset\}$ and
  $\boldsymbol{G}=\{1,\dots,q\}$. Calculate empirical estimates of
  $I(\mathcal{F}_l,X_j^{(R)})$, for $l=1,\dots,q$ and
  $j=1,\dots,p_n$, and then $I^{\mathrm{max}}_{\mathcal{F}_l,x}$, for
  $l=1,\dots,q$. Denote these estimates by
  $\hat{I}(\mathcal{F}_l,X_j^{(R)})$ and
  $\hat{I}^{\mathrm{max}}_{\mathcal{F}_l,x}$.  
\item Find $l^*=\mathrm{arg max}_{l\in\boldsymbol{G}}\{\hat{I}^{\mathrm{max}}_{\mathcal{F}_l,x}\}$. Set $\boldsymbol{S}=\mathcal{F}_{l^*}$, and $\boldsymbol{G}=\{1,\dots,q\}\setminus l^*$. Check whether or not to stop adding variables using some stopping criterion. 
\item Repeat the following steps until the stopping criterion is satisfied. 
\begin{enumerate}
\item Find $l^*=\mathrm{arg max}_{l\in
    \boldsymbol{G}}\{\hat{I}^{max}_{\mathcal{F}_l,x}-|\boldsymbol{S}|^{-1}\sum_{j\in\boldsymbol{S}}I(\mathcal{F}_l,\mathcal{F}_j)\}$. 
\item Add $\mathcal{F}_{l^*}$ to the set $\boldsymbol{S}$ and remove $l^*$ from $\boldsymbol{G}$.
\end{enumerate}
\end{enumerate}
The selected subset of variables
$\boldsymbol{\mathcal{F}}_{\boldsymbol{S}}$ are  assigned  $w_j=1/|\boldsymbol{S}|$ in the distance function, and all others are
assigned $w_j=0$. To estimate mutual information
$I(\mathcal{F}_l,X_j^{(R)})$ for discrete $\mathcal{F}_l$, we
use the empirical discrete distribution of
$(\mathcal{F}_l,X_j^{(R)})$ in the calculation. When $\mathcal{F}_l$
is continuous, a simple and effective method of estimating the mutual
information involves discretizing $\mathcal{F}_l$
\citep{ding:peng}. The marginal distribution for $\mathbf{X}^{(R)}$
depends on only the variables in $\boldsymbol{\mathcal{F}}$ that are
assigned non-zero weight, so one should include all variables that are
moderately predictive of $\mathbf{X}^{(R)}$.

Related feature selection algorithms \citep{battiti,ding:peng,estevez}
stop when a pre-specified number of features have been chosen, and use cross validation to
determine the optimal number of features.  This is impractical and
inefficient for our purposes, as the variable selection problem is not
the primary inferential or modeling focus. Rather,  it is a component of a
complex joint model that is introduced to allow for further dependence
of $\boldsymbol{\mathcal{R}}$, and especially $\mathbf{X}^{(R)}$, on
$\boldsymbol{\mathcal{F}}$. 

Instead, we make use of measures of redundancy and relevancy. 
We propose to
stop selecting variables when none remain that explain a
significant proportion of the uncertainty in any $\mathbf{X}^{(R)}$
variable, or when each remaining $\boldsymbol{\mathcal{F}}$ variable
to choose from is already well explained by the chosen set
$\boldsymbol{\mathcal{F}}_{\boldsymbol{S}}=\{\mathcal{F}_{s}:s\in
\boldsymbol{S}\}$, where each $s_j\in \{1,\dots,q\}$,
$j=1,\dots,|\boldsymbol{S}|$. 

For generic random variable $A$, let $H(A)$ be the entropy of $A$. 
Define a normalized version of mutual information,
$I^*(\mathcal{F}_l;X_{j}^{(R)})=I(\mathcal{F}_l,X_{j}^{(R)})/H(X_{j}^{(R)})$,
as the proportion of uncertainty in $X_j^{(R)}$ that is explained by
$\mathcal{F}_l$. We use this quantity to determine whether or not
${\mathcal{F}}_l$ is relevant in explaining ${X}_j^{(R)}$. Also, let
$I^*(\mathcal{F}_l;\mathcal{F}_{l'})=I(\mathcal{F}_l,\mathcal{F}_{l'})/H(\mathcal{F}_{l'})$,
for $l,l'\in 1,\dots,q$ be the proportion of uncertainty in
$\mathcal{F}_{l'}$ that is explained by $\mathcal{F}_l$.  As a measure
of redundancy between $\mathcal{F}_l$ and
$\boldsymbol{\mathcal{F}}_{\boldsymbol{S}}$, we use
$\mathrm{max}_{j\in
  \boldsymbol{S}}\{I^*(\mathcal{F}_{s_j};\mathcal{F}_l)\}$. If one or
more of the variables in $\boldsymbol{\mathcal{F}}_{\boldsymbol{S}}$
already explains a large amount of the uncertainty in $\mathcal{F}_l$,
then $\mathcal{F}_l$ is considered redundant. 

We express the relevancy stopping condition as $\mathrm{max}_{l\in
  \boldsymbol{G}}\{\mathrm{max}\{I^*(\mathcal{F}_l,X_1^{(R)}),\dots,I^*(\mathcal{F}_l,X_{q}^{(R)})\}\}<t_1$,
and the redundancy stopping condition as $\mathrm{min}_{l\in
  \boldsymbol{G}}\{\mathrm{max}_{j\in\boldsymbol{S}}\{I^*(\mathcal{F}_{s_j};\mathcal{F}_l)\}\}>t_2$.  
 Here, $t_1$ and $t_2$ are user defined thresholds between $0$ and
 $1$. A reasonable value for $t_2$ is around $0.7$ to $0.9$,
 and a reasonable value for $t_1$ is closer to $0$, such as $0.05$
 or $0.1$. Generally, information values below $0.02$ indicate a
 variable is not predictive, values $0.02$ to $0.1$ are considered
 weakly predictive, and values above $0.1$ are considered moderately
 to strongly predictive \citep{siddiqi}.
The larger $t_1$ is and the smaller $t_2$ is, the
 more parsimonious  the selected variable set is. 
Generally, the number
 of variables one uses will depend on what the variables are (in applications with knowledge about the existence of 
relationships between the $\boldsymbol{\mathcal{F}}$ and $\mathbf{X}^{(R)}$) and the importance of computational efficiency, 
as the more variables that are selected, the more time it takes the
MCMC algorithm to run.

\subsection{Posterior Inference: Full Conditionals}\label{sec:posterior}

\subsubsection*{The Mixing Parameters}
The full conditionals for the mixing parameters
$\{\boldsymbol{\theta}_h=(\boldsymbol{\beta}_h,\boldsymbol{\Sigma}_h,\boldsymbol{\psi}_h)\}$
arise by combining the 
likelihood terms in (\ref{eqn:datamodel1}) and the base distributions
in (\ref{eqn:hier_ldp}), as is standard in DP mixture models. Let $M_{h}=|\{H_i=h\}|$, or the size of cluster $h$. The
full conditional for $\boldsymbol{\beta}_h$ is matrix-normal, or
$\mathrm{vec}(\boldsymbol{\beta}_h^T)$ is multivariate normal. Let
$\boldsymbol{T}=\mathrm{diag}(\tau_1^2,\dots,\tau_{k}^2)$, and let
$\boldsymbol{D}_h$ be an $M_h$ by $k$ matrix obtained by stacking the set of row
vectors $\boldsymbol{D}(\mathbf{X}^{(R)}_i,\mathbf{Y}_i^{(F)},\mathbf{Z}_i^{(F)},\mathbf{X}_i^{(F)})$, such that $H_i=h$. Let $(\mathbf{W}^{(R)},\mathbf{Z}^{(R)})_h$ be similarly defined. The multivariate normal full conditional has covariance matrix
$\boldsymbol{V}_{\beta_h}=((\boldsymbol{T}\otimes \boldsymbol{I}_{p_o+p_c})^{-1}+(\boldsymbol{D}_h^T\boldsymbol{D}_h)\otimes
\boldsymbol{\Sigma}_h^{-1})^{-1}$, and mean vector
$\boldsymbol{V}_{\beta_h}((\boldsymbol{T}\otimes \boldsymbol{I}_{p_o+p_c})^{-1}\mathrm{vec}(\boldsymbol{\beta}_0^T)+(\boldsymbol{D}_h^T\otimes
\boldsymbol{\Sigma}_{h}^{-1})\mathrm{vec}((\mathbf{W}^{(R)})_h^T))$. 

Alternatively, we can write the distribution for $\tilde{\mathbf{W}}_{i}^{(R)}=(\mathbf{W}_{i}^{(R)},\mathbf{Z}_{i}^{(R)})$ such that $H_i=h$ as $[\tilde{\mathbf{W}}_{i,j}^{(R)}\mid \tilde{\mathbf{W}}_{i,-j}^{(R)}][\tilde{\mathbf{W}}_{i,-j}^{(R)}]$ for $j=1,\dots,p_o+p_c$. This depends on the $j$th column of $\boldsymbol{\beta}_h$, $\boldsymbol{\beta}_{h(\cdot,j)}$, only through the first univariate normal, which is
$\mathrm{N}(\tilde{\mathbf{W}}_{ij}^{(R)};\boldsymbol{D}_i\boldsymbol{\beta}_{h(\cdot,j)}+\mu^*_{i,j},\tilde{\boldsymbol{\Sigma}}_{h,j})$
 where $\mu^*_{i,j}=\boldsymbol{\Sigma}_{h(j,-j)}(\boldsymbol{\Sigma}_{h(-j,-j)})^{-1}(\tilde{\mathbf{W}}_{i,-j}^{(R)}-\boldsymbol{D}_i\boldsymbol{\beta}_{h(\cdot,-j)})$ and the conditional variance is $\tilde{\boldsymbol{\Sigma}}_{h,j}=\boldsymbol{\Sigma}_{h(j,j)}-\boldsymbol{\Sigma}_{h(j,-j)}(\boldsymbol{\Sigma}_{h(-j,-j)})^{-1}\boldsymbol{\Sigma}_{h(-j,j)}$. Here $\boldsymbol{D}_i=\boldsymbol{D}(\mathbf{X}^{(R)}_i,\mathbf{Y}_i^{(F)},\mathbf{Z}_i^{(F)},\mathbf{X}_i^{(F)})$.

We therefore can update each column $j=1,\dots,p_o+p_c$ of the matrix $\boldsymbol{\beta}_h$ from a multivariate normal with variance $\boldsymbol{V}_{{h,j}}=(\boldsymbol{T}^{-1}+\tilde{\boldsymbol{\Sigma}}_{h,j}^{-1}(\boldsymbol{D}_h^T\boldsymbol{D}_h))^{-1}$ and mean $\boldsymbol{V}_{{h,j}}(\boldsymbol{T}^{-1}\boldsymbol{\beta}_{0\cdot j}+\tilde{\Sigma}_{h,j}^{-1}\boldsymbol{D}_h^T((\tilde{\mathbf{W}^{(R)}})_{h(\cdot,j)}-\mu^*_{h,j})$ where $\tilde{\Sigma}_{h,j}$ is the conditional variance of $(\tilde{\mathbf{W}}_{ij}^{(R)}\mid \tilde{\mathbf{W}}_{i,-j}^{(R)},H_i=h)$ and $\mu^*_{h,j}$ is obtained by concatenating the set of $\mu^*_{i,j}$ over $\{i:H_i=h\}$. 

When cluster $h$ is empty, i.e., $M_h=0$,  we draw from the base distribution, which we can do by simulating $\mathrm{vec}(\boldsymbol{\beta}_h) \sim \mathrm{N}(\mathrm{vec}(\boldsymbol{\beta}_0),\boldsymbol{I}_{p_o+p_c}\otimes\boldsymbol{T})$. 

The covariance matrices
$\boldsymbol{\Sigma}_h$ are updated from
$\mathrm{IW}(\nu+M_{h},\boldsymbol{S}+\sum_{\{i:H_i=h\}}(\tilde{\mathbf{W}}_i^{(R)}-\boldsymbol{\beta}_{h}\boldsymbol{D}_i)(\tilde{\mathbf{W}}_i^{(R)}-\boldsymbol{\beta}_{h}\boldsymbol{D}_i)^T)$
where
$\boldsymbol{D}_i=\boldsymbol{D}(\mathbf{X}^{(R)}_i,\mathbf{Y}_i^{(F)},\mathbf{Z}_i^{(F)},\mathbf{X}_i^{(F)})$. If
$M_h=0$, $\boldsymbol{\Sigma}_h$ is updated from the base
distribution $\mathrm{IW}(\nu,\boldsymbol{S})$. The probability
vectors of the categorical distributions have full conditionals which
are Dirichlet distributed: $\boldsymbol{\psi}_h^{(j)}
\sim \mathrm{Dirichlet}(a_1+\sum_{\{i:H_{i}=h\}}
1(X_{ij}^{(R)}=1),\dots,a_{d_j}+\sum_{\{i:H_i=h\}}
1(X_{ij}^{(R)}=d_j))$. Again, if $M_h=0$, then the update is
simply $\mathrm{Dirichlet}(a_1,\dots,a_{d_j})$. 

We simulate from the posterior distribution of $\boldsymbol{\beta}_h$, $\boldsymbol{\Sigma}_h$, and $\boldsymbol{\psi}_h^{(j)}$, for $h=1,\dots,N$, and $j=1,\dots,p_n$ using the full conditionals given above.

\subsubsection*{The Mixture Allocation Variables and Local DP Parameters}

The mixture configuration variables $H_i$, where $i=1,\dots,n$ are
simulated from categorical distributions.  We have 
\[p(H_i\mid \dots)\propto \sum_{l=1}^{N(\boldsymbol{f}_i)}{p_l}'(\boldsymbol{f}_i)\delta_{\pi_l(\boldsymbol{f}_i)}(\cdot)\],
where ${p_l}'(\boldsymbol{f}_i)\propto p_l(\boldsymbol{f}_i)\mathrm{N}(\boldsymbol{\beta}_{\pi_l(\boldsymbol{f}_i)}\boldsymbol{D}(\mathbf{X}^{(R)}_i,\mathbf{Y}_i^{(F)},\mathbf{Z}_i^{(F)},\mathbf{X}_i^{(F)}),\boldsymbol{\Sigma}_{\pi_l(\boldsymbol{f}_i)})\prod_{j=1}^{p_n}\psi^{(j)}_{\pi_l(\boldsymbol{f}_i),X_{ij}^{(R)}}$.

The full conditionals for $V_h$, $h=1,\dots,N$, are derived as follows:
\begin{eqnarray} 
p(V_h\mid \dots)&\propto& \mathrm{beta}(V_h;1,\alpha)\prod_{i=1}^n \sum_{l=1}^{N(\boldsymbol{C}_i)}p_l(\boldsymbol{f}_i)\delta_{\pi_l(\boldsymbol{f}_i)}(H_i)\nonumber \\
&\propto& \mathrm{beta}(V_h;1,\alpha)\prod_{\substack{i:H_i=h \\ H_i \neq \pi_{N(c_i)}(c_i)}}V_h \prod_{\substack{i: \, H_i>h\\ h\in {\eta}_{\boldsymbol{f}_i}}} (1-V_h) \nonumber \\
&\propto& \mathrm{beta}(V_h;1+\sum_{i=1}^n 1(H_i=h, \, H_i\neq \pi_{N(\boldsymbol{f}_i)}(\boldsymbol{f}_i)),\alpha+\sum_{i=1}^n 1(H_i>h \text{ and } h\in\eta_{\boldsymbol{f}_i})).\nonumber
\end{eqnarray}

The full conditional for $\boldsymbol{\Gamma}_h$ is 
\begin{equation*}
p(\boldsymbol{\Gamma}_h\mid\dots)\propto p(\boldsymbol{\Gamma}_h)\prod_{\{i:H_i=h\}}1(d(
\boldsymbol{f}_i,\boldsymbol{\Gamma}_h)<d^*).
\end{equation*}
The full conditional in general depends on the choice of distance
function, which in CMM-Mix is  $d(
\boldsymbol{f}_i,\boldsymbol{\Gamma}_h)=\sum_{j=1}^qw_jd_j^e(f_{ij},{\Gamma}_{hj})$. The term $\sum_{j=1}^qw_jd_j^e(f_{ij},{\Gamma}_{hj})<d^*$ implies $a_{il}<\Gamma_{hl}<b_{il}$, for any $l=1,\dots,q$. Combined with the product of independent uniforms prior for $\boldsymbol{\Gamma}_h$, the  full conditionals for each $\Gamma_{hl}$ are uniform. 

For $l=1,\dots,q_o$, $\Gamma_{hl}$ corresponds to ordinal variable $\mathcal{F}_l$, with $\mathcal{F}_l\in\{1,\dots,k_l^{(F)}\}$, and $d(\boldsymbol{\mathcal{F}}_{i},\Gamma_{h})<d^*$ implies 
$a_{il}<\Gamma_{hl}<b_{il}$ where $a_{il}=\mathcal{F}_{il}+((k_l^{(F)}-1)/w_l)(d^*-\sum_{j\neq l}w_jd_j^{e}(\mathcal{F}_{ij},\Gamma_{hj}))$ and $b_{il}=\mathcal{F}_{il}-((k_l^{(F)}-1)/w_l)(d^*-\sum_{j\neq l}w_jd_j^{e}(\mathcal{F}_{ij},\Gamma_{hj}))$.
Therefore the full conditional for $\Gamma_{hl}$ is uniform, with lower bound equal to the smallest value in $\{1,\dots,k_l^{(F)}\}$ that is greater than max$_{\{i:H_i=h\}}\{a_{il}\}$, and upper bound equal to the largest value in $\{1,\dots,k_l^{(F)}\}$ that is less than min$_{\{i:H_i=h\}}\{b_{il}\}$. For $\{l:w_l=0\}$, $\Gamma_{hl}$ is simulated from the prior, which is uniform on $\{1,\dots,k_l^{(F)}\}$.

For $l=q_o+1,\dots,q_o+q_c$, $\Gamma_{hl}$ corresponds to continuous variable $\mathcal{F}_l$, and under a uniform$(a_l^{\Gamma},b_l^{\Gamma})$ prior for $\Gamma_{hl}$, $d(\boldsymbol{\mathcal{F}}_{i},\Gamma_{h})<d^*$ implies $a_{il}<\Gamma_{hl}<b_{il}$ where $a_{il}=\mathcal{F}_{il}+(\mathrm{max}(\mathcal{F}_{l})-\mathrm{min}(\mathcal{F}_l))(d^*-\sum_{j\neq l}w_jd_j^{e}(\mathcal{F}_{ij},\Gamma_{hj}))/w_l$ and $b_{il}=\mathcal{F}_{il}-(\mathrm{max}(\mathcal{F}_{l})-\mathrm{min}(\mathcal{F}_l))(d^*-\sum_{j\neq l}w_jd_j^{e}(\mathcal{F}_{ij},\Gamma_{hj}))/w_l$.
Therefore the full conditional for $\Gamma_{hl}$ is uniform, with lower bound equal to $\mathrm{max}(\mathrm{max}_{\{i:H_i=h\}}\{a_{il}\},a_{l}^{\Gamma})$, and upper bound equal to $\mathrm{min}(\mathrm{min}_{\{i:H_i=h\}}\{b_{il}\},b_{l}^{\Gamma})$.   For $\{l:w_l=0\}$, $\Gamma_{hl}$ is simulated from the prior, which is uniform$(a_{l}^{\Gamma},b_{l}^{\Gamma})$.

For $l=q_o+q_c+1,\dots,q$, $\Gamma_{hl}$ corresponds to nominal variable $\mathcal{F}_l\in\{1,\dots,d_l^{(F)}\}$, and $d(\boldsymbol{\mathcal{F}}_{i},\Gamma_{h})<d^*$ implies 
\[1(\mathcal{F}_{il}\neq \Gamma_{hl})<b_{il}=\left(d^*-\sum_{j\neq l}w_jd_j^e(\mathcal{F}_{ij},\Gamma_{hj})\right)/w_l\]
and therefore if $\mathrm{min}_{\{i:H_i=h\}}(b_{il})<1$, this implies a degenerate full conditional for $\Gamma_{hk}$ at $\mathcal{F}_{il}$. Otherwise, $\Gamma_{hl}$ is drawn randomly from $\{1,\dots,d_{l}^{(F)}\}$. For $\{k:w_k=0\}$, $\Gamma_{hl}$ is simulated from the prior, which is uniform on $\{1,\dots,d_{l}^{(F)}\}$. 

\subsubsection*{The Base Distribution Hyperparameters}

The full conditional for $\alpha$ is the same as under the standard DP applied with truncation. Under the prior $\alpha\sim \mathrm{gamma}(a_\alpha,b_\alpha)$, then the posterior for $\alpha$ is proportional to $\mathrm{gamma}(a_\alpha+N,b_\alpha-\sum_{h=1}^{N} \log(1-V_h))$.

Under the prior $\boldsymbol{\beta}_0\sim \mathrm{MN}_{k\times (p_o+p_c)}(\boldsymbol{0},\boldsymbol{I},h\boldsymbol{I})$, the full conditional for each $\boldsymbol{\beta}_{0mr}$, $m=1,\dots,k$, $r=1,\dots,p_o+p_c$ is then $\mathrm{N}((h^{-1}+N\tau_m^{-2})^{-1}(\tau_m^{-2}\sum_{l=1}^N\beta_{l(m,r)}),(h^{-1}+N\tau_m^{-2})^{-1})$. 

Assuming $\tau^2_j\sim \mathrm{IG}(a_\tau,b_\tau)$ for $j=1,\dots,k$, we obtain a posterior full conditional for $\tau^2_j$ which is $\mathrm{IG}(a_\tau+0.5N(p_o+p_c), b_\tau+0.5\sum_{h=1}^N\sum_{j=1}^{p_o+p_c}({\beta}_{h(l,j)}-{\beta}_{0(l,j)})^2)$.

With a prior $\boldsymbol{S}\sim \mathrm{Wish}(a_S,\boldsymbol{B}_S)$, $\boldsymbol{S}$ is sampled from Wish$(N\nu+a_S, (\boldsymbol{B}_S^{-1}+\sum_{h=1}^N \boldsymbol{\Sigma}_h^{-1})^{-1})$.

\subsubsection*{Latent Continuous Variables and Missing Data Imputation}
The latent continuous random variables $W_{ij}^{(R)}$, $i=1,\dots,n$, and $j=1,\dots,p_o$ have full conditionals $p(W_{ij}^{(R)}\mid \dots)\propto \mathrm{N}(W_{ij}^{(R)};\tilde{\mu}_{ij},\tilde{\sigma}^2_{ij})$
where $\tilde{\mu}_{ij}$ and $\tilde{\sigma}^2_{ij}$ are the conditional normal mean and variance for $W_{ij}^{(R)}$ resulting from the multivariate normal distribution \\
$\mathrm{N}(\mathbf{W}_i^{(R)},\mathbf{Z}_i^{(R)};\boldsymbol{D}(\mathbf{X}^{(R)}_i,\mathbf{Y}_i^{(F)},\mathbf{Z}_i^{(F)},\mathbf{X}_i^{(F)})\boldsymbol{\beta}_{H_i},\boldsymbol{\Sigma}_{H_i})$. If $Y_{ij}^{(R)}$ is observed, then $W_{ij}^{(R)}$ must lie in the interval $(\gamma_{j,Y_{ij}^{(R)}-1},\gamma_{j,Y_{ij}^{(R)}}]$, i.e., its full conditional is truncated normal.
If $Y_{ij}^{(R)}$ is missing, then $W_{ij}^{(R)}$ is sampled from the normal distribution without any truncation. 

Similarly, missing $Z_{ij}^{(R)}$ are simulated from the normal distribution $\mathrm{N}(Z_{ij}^{(R)};\tilde{\mu}_{ij},\tilde{\sigma}^2_{ij})$ where $\tilde{\mu}_{ij}$ and $\tilde{\sigma}^2_{ij}$ are the conditional normal mean and variance for $Z_{ij}^{(R)}$ resulting from the multivariate normal $\mathrm{N}((\mathbf{W}_i^{(R)},\mathbf{Z}_i^{(R)});\boldsymbol{D}(\mathbf{X}^{(R)}_i,\mathbf{Y}_i^{(F)},\mathbf{Z}_i^{(F)},\mathbf{X}_i^{(F)})\boldsymbol{\beta}_{H_i},\boldsymbol{\Sigma}_{H_i})$.

In order to improve mixing, we can block update the $W_{ij}^{(R)}$ and $Z_{ij}^{(R)}$ using the fact that $p(\tilde{\mathbf{W}}_i^{(R)}\mid \mathrm{data},\theta)=p(\tilde{W}_{i1}^{(R)}\mid \mathrm{data},\theta)\prod_{j=2}^{p_o+p_c}p(\tilde{W}_{ij}^{(R)}\mid (\tilde{W}_{i1}^{(R)},\dots,\tilde{W}_{i,j-1}^{(R)}),\mathrm{data},\theta)$, where $\theta$ indicates all model parameters. 
Letting $\boldsymbol{D}_{i}=\boldsymbol{D}(\mathbf{X}^{(R)}_i,\mathbf{Y}_i^{(F)},\mathbf{Z}_i^{(F)},\mathbf{X}_i^{(F)})$, we sample $\tilde{W}_{i1}^{(R)}\sim \mathrm{N}(\tilde{W}_{i1}^{(R)};\boldsymbol{D}_{i}\boldsymbol{\beta}_{H_i(\cdot,1)},(\Sigma_{H_i})_{11})$, with truncation to $(\gamma_{1,Y_{i1}^{(R)}-1},\gamma_{1,Y_{i1}^{(R)}}]$ if $Y_{i1}^{(R)}$ is observed. For $j=2,\dots,p_o+p_c$, let $\mu^*_{ij}$ and $(\sigma^*)^2_{ij}$ denote the conditional mean and variance for $(\tilde{W}_{ij}^{(R)}\mid \tilde{W}_{i1}^{(R)},\dots,\tilde{W}_{i,j-1}^{(R)}$), and simulate $\tilde{W}_{ij}^{(R)}\sim \mathrm{N}(\tilde{W}_{ij}^{(R)};\mu^*_{ij},(\sigma^*)^2_{ij})$, with truncation to  $(\gamma_{j,Y_{ij}^{(R)}-1},\gamma_{j,Y_{ij}^{(R)}}]$ for $j=2,\dots,p_o$ if $Y_{ij}^{(R)}$ is observed.

Missing $X_{ij}^{(R)}$ are simulated from categorical distributions on $\{1,\dots,d_{j}^{(R)}\}$ with probabilities proportional to 
\[\left(\psi^{(j)}_{H_i,1}\mathrm{N}(\mathbf{W}_i^{(R)},\mathbf{Z}_i^{(R)};\boldsymbol{D}_1\boldsymbol{\beta}_{H_i},\boldsymbol{\Sigma}_{H_i}),\dots,\psi^{(j)}_{H_i,d_j^{(R)}}\mathrm{N}(\mathbf{W}_i^{(R)},\mathbf{Z}_i^{(R)};\boldsymbol{D}_{d_j^{(R)}}\boldsymbol{\beta}_{H_i},\boldsymbol{\Sigma}_{H_i})\right)\]
where $\boldsymbol{D}_l$ indicates the current value of $\boldsymbol{D}(\mathbf{X}^{(R)}_{i},\mathbf{Y}_i^{(F)},\mathbf{Z}_i^{(F)},\mathbf{X}_i^{(F)})$ in which $X_{ij}^{(R)}=l$.

\end{document}